\documentstyle[aps,pre,multicol,psfig]{revtex}

\newcommand{\beq}{\begin{equation}}
\newcommand{\eeq}{\end{equation}}
\newcommand{\bea}{\begin{eqnarray}}
\newcommand{\eea}{\end{eqnarray}}
\newcommand{\bml}{\begin{mathletters}}
\newcommand{\eml}{\end{mathletters}}

\input epsf

\begin{document}
\draft

\title{Model for the hydration of non-polar compounds and polymers}

\author{Pierpaolo Bruscolini\cite{P}} 

\address{Istituto Nazionale per la Fisica della Materia (INFM) and
Dipartimento di Fisica, Politecnico di Torino, \\
Corso Duca degli Abruzzi 24, I-10129 Torino, Italy} 

\author{Lapo Casetti\cite{lapo}}

\address{Istituto Nazionale per la Fisica della Materia (INFM), Unit\`a di
Ricerca di Firenze, \\ Largo Enrico Fermi 2, I-50125 Firenze, Italy} 

\date{\today}

\maketitle

\begin{abstract}
We introduce an exactly solvable statistical-mechanical model of the 
hydration of non-polar compounds, based on grouping water molecules 
in clusters where hydrogen bonds and isotropic interactions occur; 
interactions between clusters are neglected.
Analytical results show that an effective  strengthening of hydrogen bonds
in the presence of the solute, together with a geometric reorganization  
of water molecules, are enough to yield hydrophobic behavior. 
We extend our model to describe a non-polar homopolymer
in aqueous solution, obtaining a clear evidence of both ``cold'' and ``warm''
swelling transitions. This suggests that our model could be relevant to
describe some features of protein folding.
\end{abstract}

\begin{multicols}{2}   
\narrowtext  
\section{Introduction}
 The physical properties of liquid water, despite the extensive studies
devoted to their investigation, are not yet completely understood. The
behavior of water when nonpolar solutes are present is still a matter
of debate, even if simplified models \cite{Muller,KoWi} as well as
numerical simulations (see Ref.~\cite{Dillea99} and references quoted
therein) have shed some light on it.  The hydration of non-polar
compounds has a free energy cost, resulting in a net ``attraction''
among solute molecules which tend to aggregate in order to minimize
exposition to water (``hydrophobic effect'').  More precisely, upon
hydration of a single molecule from the gaseous phase, the free energy
and specific heat changes $\Delta F$ and $\Delta C$ are positive over
the entire temperature range accessible to experiments; the internal
energy end entropy differences $\Delta E$ and $\Delta S$ are negative
for low temperatures, then become positive when the temperature is
raised. The two temperatures at which $\Delta E$ and $\Delta S$ are
zero are commonly referred to as $T_H$ and $T_S$, respectively.  The
origin of the lowering of the entropy for $T < T_S$ has been related
to the ordering of water molecules around the solute to preserve bonds
\cite{Creibook,Dillea98} as well as to the opening of a ``cavity'' in
water to settle the solute, without ordering of water molecules
\cite{ArtHay98}; but the question is still controversial.
Hydrophobicity plays a fundamental role in many physical processes,
among which the long-standing problem of protein folding, since the
hydrophobic effect is believed to be one of the most relevant
``forces'' guiding the protein into the folded native state, i.e., the
unique spatial structure in which the protein is biologically active
\cite{Dill}. At present, the whole folding process can be studied only
using coarse-grained
models (all-atom simulations of proteins and solvent molecules are
feasible only on timescales several order of magnitude smaller than
the folding time).  In many models water is not explicitly considered
and solvent-induced aggregation is described through
temperature-independent effective attraction among nonpolar monomers.
An important feature of real (globular) proteins is that the native
state becomes unstable not only at high temperature ({\em warm
unfolding}), but also at low temperatures ({\em cold unfolding})
\cite{Privalov}; cold swelling has also been detected in some
homopolymers \cite{pnipam,Wangea98}.  
A major drawback of the above simplified
models is that they cannot account for cold unfolding, because there
the native (folded) state is identified with the model ground state,
so that the protein is folded at low temperatures down to $T=0$.

In this paper we propose a ``minimal'' model capable of reproducing
the thermodynamic hallmarks of hydrophobicity in single molecule's
hydration, within the framework of the ``water ordering'' hypothesis;
moreover, we keep it as simple as to be analytically integrable, in
order to apply it to polymer hydration.  
For its simplicity, the model is not intended to reproduce all the
details of hydrophobic behavior: rather  it 
gives a  correct overall description of the phenomenology and 
allows to understand what are the key ingredients that produce it.

In the case of a nonpolar
homopolymer we recover both ``cold'' and ``warm'' swelling
transitions, thus strengthening the idea, already put forward in
Refs.~\cite{pl2000}, that an explicit, though simplified, description
of water can also provide a framework for a unified treatment of both
transitions in proteins.

\section{The model}

\label{sec_model}
We describe water molecules as two-dimensional objects with three 
hydrogen-bonding  arms, namely, the planar projection of the
tetrahedral coordination of water molecules, as in
Refs.~\cite{BellBenNaim,Dillea98}.
Simulations show that such a crude model
reproduces the  main features of water \protect\cite{Dillea98}.
The molecules in contact with a solute will be
referred to as  the {\em hydration shell}. Hydrophobicity is a collective
phenomenon, yet experiments \cite{Privalov}  show additivity in the 
nonpolar area exposed to the solvent. Hence, there  exists a cluster scale
above which correlations may be neglected. We consider the  hydration shell to
be made up of $z$ such clusters, each of $m$ interacting molecules, and we
disregard interactions between clusters. We assume that the ground state of a
cluster is characterized by a completely formed  hydrogen-bond network,  both
in  the bulk and in the hydration shell case. 
Thus each molecule in the ground state has  on average 
$3/2$ hydrogen bonds. 
 Hydrogen-bond energies are very sensitive to the
geometry of the bond, and the latter will be affected by the presence
of a solute,  so that  
we assume that bond energies may differ in the
``bulk'' and ``shell''
cases. We model this by simply assuming that there is an energy difference
(per molecule) $K$ between bulk and shell ground states, given by
\beq
K = \frac{3}{2} \left(h_b - h_s \right) + J ~,
\label{eq:defk}
\eeq
where $h_b$ and $h_s$ are the hydrogen-bond energies for bulk and shell 
molecules, respectively. The fact that $h_b \not = h_s$ is not, in
principle, the unique possible cause of an energy shift in the ground
states: not only a solute perturbs the water hydrogen bonds, but also has
a different direct interaction with water. 
To take into account this kind of  contributions to the ground-state
energy difference, assuming that they
can be  reasonably averaged on the molecular scale, we introduce the
constant $J$ in Eq.\ (\ref{eq:defk}) above.

\subsection{Partition function of the hydration shell}

\begin{figure}
\epsfysize= 4 truecm 
\begin{center}
\mbox{\epsfbox{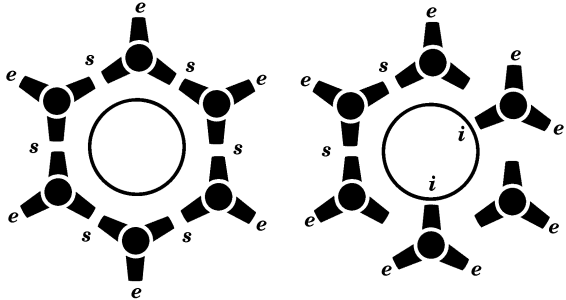}} 
\end{center}
\caption{Cluster of water molecules. Bonds may be formed
or broken, according to energy. The circle
represents a solute (shell case) or (an)other water molecule(s) (bulk).
{\em Left}: ordered arrangement  
with molecules in state
``E'' (one ``external'' ($e$) and two ``on-shell'' ($s$) arms). When all bonds
are formed this is the ground state.
{\em Right}: in state-I molecules have two
$e$ arms and an internal ($i$) one,
which cannot form a bond if a solute is present. }
\label{fig:model}
\end{figure}

Our goal is to model the thermodynamics of hydration, hence  we
need to  evaluate the partition function for the cluster in both
cases: 
\beq
{\cal Z}_{\bullet}^{\text{clu}} = 
\int_0^\infty d\varepsilon \, g_{\bullet} (\varepsilon)\,
e^{-\beta {\cal H}_\bullet^{\text{clu}} ({\varepsilon})} \, ,
\label{zclu}
\eeq
where $\bullet=b,s$  (bulk or shell, respectively). Here,
\beq
{\cal H}_\bullet^{\text{clu}} = 
  \varepsilon + K m \,
\delta_{\bullet, s} ~,
\label{hamclu}
\eeq
and $g_{\bullet} (\varepsilon)$ is the density of states of 
the cluster at
the energy $\varepsilon$ above its  ground state:  it encodes
all the important features determining the
system thermodynamics.
How can we estimate $g_{\bullet} (\varepsilon)$? 
Given an energy
$\varepsilon$ to the cluster we expect to have a certain number of
broken bonds and  assume equipartition of the energy 
on the various degrees of freedom
of the cluster. Moreover, bond breaking allows molecules  to 
access a wider configuration space: while the ground state is unique,
different geometrical arrangements of the molecules can  have the same
number of bonds.
 To describe it simply enough, 
we take each bond to be represented by
a harmonic potential, and  to  be broken independently of the others.  
Thus, at the cluster energy $\varepsilon$ there will be a probability
\beq
p_{\bullet}(\lambda,\varepsilon) = {{\frac{3}{2} m} \choose {\lambda}}\,
p_{1 \bullet}(\varepsilon)^\lambda \, \left( 1 - p_{1
\bullet}(\varepsilon) 
\right)^{\frac{3}{2} m - \lambda} 
\label{plam}
\eeq
to break $\lambda$ bonds. 
Here 
$p_{1 \bullet}(\varepsilon)$ is the probability 
that a bond acquires an energy larger
than  $h_{\bullet}$ and breaks, when  $\varepsilon$ is
equipartitioned  on $D$ degrees of freedom: 
$p_{1 \bullet}(\varepsilon) = e^{-h_{\bullet}/\tilde{T}}$ and  
$\tilde{T} = 2 \varepsilon / D $. For the sake of simplicity 
we assume $D= 2 m f$  ($f$ are the degrees of freedom of one molecule:
$f=3$ in two dimensions) as if molecules were at least singly  
bonded  at all energies. In the same spirit we take the density of
states $\omega(\lambda,\varepsilon)$ of the system
of harmonic oscillators, resulting when $\lambda$ bonds are  broken, to
be independent of  $\lambda$: 
$\omega(\lambda,\varepsilon) \simeq \omega(\varepsilon) =
{\cal C} \, \varepsilon^{\frac{D}{2} - 1}$,
where ${\cal C}$ is a constant. 
To account for the geometric degeneracy we assume just two orientations
for each molecule,  (E and I; see Fig.~\protect\ref{fig:model}), and give
an estimate of $\gamma_{\bullet} \left( \nu \right)$, 
the number of accessible geometric
arrangements of the cluster when $\lambda= 3 m/2-  \nu$  bonds are broken.
Then, our ansatz for $g_{\bullet} (\varepsilon)$ will be the sum, over
$\lambda$, of the probability $p_{\bullet}(\lambda,\varepsilon)$ of breaking
$\lambda$ bonds, times the density of states $\omega(\varepsilon)$ of
each geometric configuration, times the number $\gamma_{\bullet} (\nu)$ of 
configurations at $\lambda$ broken bonds:
\beq
g_{\bullet}(\varepsilon) = 
\sum_{\lambda = 0}^{\frac{3}{2}m} p_{\bullet}(\lambda,\varepsilon)\,
\omega(\varepsilon) \, 
\gamma_{\bullet} \left(\frac{3 m}{2} - \lambda\right)~. 
\label{dos}
\eeq
In the estimation of $\gamma_{\bullet}(  \nu )$ the bookkeeping 
is easier if we
attribute each bond to just one of the two bonding molecules. 
Each cluster can have no more than $3 m/2$  bonds: to ensure this,
we take $m$ even and fictitously split the cluster in two equal sets, 
{\em a} and  {\em b}.
Group {\em a}-molecules will form at most one bond: 
a shell one when in state-E 
and an external one when in state-I.
Group {\em b}-molecules  form at most two bonds:
the external and a
shell one in state-E, 
an external and the internal one in state-I. 
Internal bonds will be  possible only for  
({\em b},I) molecules, if they find a bonding partner.
The number of internal bonding sites, $q_\bullet$,
will distinguish ``shell'' from ``bulk''  cases: in the former
$q_s=0$, while in the latter $0< q_b \le m$. In fact, 
due to the geometry of  the internal  
water molecules, possibly not all 
the $m$ positions allow internal bonds, 
even in the bulk case.
State-E  molecules can form one shell bond, while state-I cannot:
hence, the probability of a configuration with $s$ state-I molecules 
separating $(m-s)$ state-E ones, 
so that $k$ shell bonds are forbidden and 
$\chi = m-s-k$ can be formed, is \cite{note_gamma}
\beq
\pi (s, k) = 2^{-m} {{s}\choose{k}} \, {{m-s} \choose {k}}.
\label{pi_plain}
\eeq
The probability 
that one of these configurations also has  
$i$ internal bonds depends, first, on the probability of 
fishing out $j$ ({\em b},I)-molecules among the $s$ in state-I, 
given the total number of molecules $m$ and of $b$-molecules 
$m/2$: $\pi_{\text{h}} (j; s,m/2,m)$, where
$\pi_{\text{h}}$ is a hypergeometric probability, i.e.,
\beq
\pi_{\text{h}} (s; n, S, N) = {{S} \choose {s}}\, {{N-S} \choose
{n-s}}\, \left[ {{N} \choose {n}} \right]^{-1} \,.  
\label{pi_h}
\eeq
Then, one must consider the
probability of placing $i$ of the $j$ 
molecules with internal arms, in the $q_\bullet$ good positions 
for bond formations, on a total of $m$ possibilities: 
$\pi_{\text{h}} (i; j, q_\bullet, m)$. 
The product of all the above probabilities gives the fraction 
of conformations  that has  $s$ state-I molecules and
is able to form $\chi$ shell bonds, $i$ internal ones and $m/2 +s-j$ 
external ones 
(assuming that all the external arms form 
bonds). 
If we now let the bonds be also broken, 
we must notice that
any pattern geometrically allowing more than $\nu$ bonds will also 
contribute to $\gamma_{\bullet} ( \nu)$: we just need to consider 
the excess bonds as broken -- we choose them among the 
external, internal and 
shell ones with a binomial probability 
$\pi_{\text{b}} \left(n, N  \right) \equiv
\pi_{\text{b}} \left(n, N, 1/2 \right)$, where 
\beq
\pi_{\text{b}} (n, N,p) = {{N} \choose {n}} \,p^n (1-p)^{(N-n)} \,.
\label{pi_b}
\eeq
 Hence,  upon summing over all 
the geometric arrangements that can contribute to a 
pattern with $\nu$ bonds, we finally get
\beq
\gamma_{\bullet} \! \left( \nu \right) \!\!=\!\!\! 
\sum_{\nu_s=0}^{\nu} \sum_{\nu_e=0}^{\nu-\nu_s}  
\! \sum_{s=0}^{m}\! \sum_{k=0}^{k_{\text{max}}} \! \sum_{j=0}^{s} 
\! \sum_{i=0}^{j}   
\xi_\bullet \! \left( s, k, j, i,  \nu_e, \nu_s,\nu_i \right) 
, 
\label{gamma}
\eeq
where  $\nu_i = \nu- \nu_s- \nu_e$,
$k_{\text{max}}=\min(s, m-s-\nu_s)$ and
\bea
\lefteqn{\!\!\!\!\!\!
\xi_\bullet \left( s, k, j, i, \nu_e,\nu_s,\nu_i \right)=  
\pi (s, k) \pi_{\text{h}} \left(j; s,\frac{m}{2},m \right) 
\pi_{\text{h}} (i; j, q_\bullet, m)}\nonumber \\ 
&&
\,\times \pi_{\text{b}} \left(\nu_i, i \right)  
\pi_{\text{b}} \left(\nu_e, \frac{m}{2}+s-j \right)
\pi_{\text{b}} \left(\nu_s, m-s-k \right)\,.
\label{xi}
\eea
 The partition function can now be evaluated from Eq.\ (\ref{zclu}) and
Eqs.\ (\ref{dos},\ref{plam},\ref{gamma}). We get: 
\beq
{\cal Z}_{\bullet}^{\text{clu}} = 
A_\bullet \, e^{-\beta K m  \delta_{\bullet,s}}
\label{zclu2}
\eeq
where
$
A_\bullet ={\cal C} 
\sum_{\lambda = 0}^{\frac{3}{2} m} {{\frac{3}{2} m} \choose {\lambda}}\,
\gamma_{\bullet} \left(3 m/ 2 - \lambda\right) 
I_\bullet \left( \lambda \right)\, ,
$
${\cal C}$ is a constant and 
\beq
I_\bullet ( \lambda ) = 
\frac{\left(  \frac{D}{2} \!- \! 1 \right)!}{ 
\beta^{\frac{D}{2}}} \delta_{\lambda,0} +  
2 \!\!\!\! \sum_{j=\delta_{\lambda,0}}^{\lambda'} \!\!\! 
(-1)^j  {{\lambda'} \choose {j }}\, 
\sigma_\bullet^{\frac{D}{4}} K_{\frac{D}{2}}(\tau_\bullet) \,,  
\eeq
with $K_D(\tau_\bullet)$ the Bessel-$K$ function, 
$\lambda'= 3 m/2 -\lambda$,  and 
\beq
\sigma_{\bullet}=(2 \beta)^{-1} D  h_\bullet \left(j+\lambda \right)
\, , ~
\tau_{\bullet}= \sqrt{2  \beta h_\bullet D \left(j+\lambda \right)}\,.
\eeq

\subsection{Statistical mechanics of non-polar hydration}

\begin{figure}
\epsfysize= 4 truecm 
\begin{center}
\mbox{\epsfbox{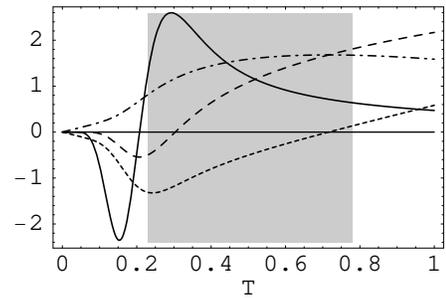}} 
\end{center}
\caption{Free energy $\Delta F$ (dash-dotted), 
energy $\Delta E$ (dashed), entropy $T \Delta S$ (dotted)
and specific heat $\Delta C/3$ (solid line; $1/3$ is for rendering purposes)
changes upon hydration of a nonpolar solute.
Here $z=4$, $m=4$, $q_s=0$, $q_b=m/2$, $h_s/h_b =1.2$, $J=0.3$
(so that $K=0$). Boltzmann's constant $k_B = 1$
so that the specific heat is adimensional. Energies and temperatures  are 
in units of $h_b$. The shaded region can be compared
with experiments \protect\cite{Privalov} and simulations
\protect\cite{Dillea98}.}
\label{fig:solutetherm}
\end{figure}

We now consider the transfer of a nonpolar 
molecule from its gas phase to water (i.e., hydrophobic hydration). 
We describe it by taking $z$  clusters, of $m$ water molecules,  
in the bulk case and substituting the internal water  with the
solute. The free energy change $\Delta F$ is given by
\beq
\beta \Delta F = 
- z \log{x}\,,
\label{formulazze}
\eeq
with  $x=  A_{s} A_{b}^{-1}\,
\exp (- \beta K m )$. 
Similar expressions hold for  energy, entropy and specific heat changes. 
The temperature dependence of these functions, 
reported in Fig.\ \ref{fig:solutetherm}, shows the hallmarks 
of hydrophobic  behavior:    
we find a maximum in the free energy
cost and minima  in both $\Delta E$ and $T \Delta S$; 
as $T$ grows, first
$\Delta E$ and then 
$T \Delta S$ cross the zero -- at $T_H$ and $T_S$, respectively -- 
and eventually become positive. 
Moreover, we correctly find a
pronounced and positive peak in the specific heat difference $\Delta
C$. This cannot be found within the information theory
approximation to the cavity model: see Ref.~\cite{ArtHay99}.
 It turns out that both a ratio $h_s/h_b >1$ and
the difference between  $\gamma_{b} (\nu)$ and  $\gamma_{s} (\nu)$
concur to produce the correct thermodynamic  behavior: 
with $h_s= h_b$ the difference between
$\gamma_b (\nu)$ and $\gamma_s (\nu)$ is not sufficient to
produce the minimum in $\Delta E$, while  $\gamma_{b} ( \nu)=  
\gamma_{s} ( \nu)$ (i.e.,  same chance of internal bonds, 
$q_{b} =  q_{s}$) yields too small a peak in $\Delta C$. 

Then, according to our model, hydrophobicity appears as 
related to the average strengthening of hydrogen bonds, together with
geometric pattern changes, upon the introduction of a nonpolar solute.

In particular,  having tighter shell
bonds than bulk ones ($h_s>h_b$) appears to be essential, 
in our model, to get a negative
minimum in $\Delta E$. Indeed, even starting from equal energy ground
states in the bulk and shell cases (i.e., $K=0$), if $h_s>h_b$
 it will be easier to break bonds in bulk than in shell, according to
Eq.~(\ref{plam}). This yields that at ``intermediate'' temperatures shell
water will have lower energy than bulk, i.e. $\Delta E<0$.

The predictions on the trends of thermodynamic functions  are
in qualitative agreement with experimental results for noble
gases (see the results by Crovetto and coworkers \cite{Croea} as
reported in \cite{Dillea98}) and for nonpolar surfaces in proteins
\cite{Privalov}.  They are also in agreement with the Monte Carlo 
results for the two-dimensional ``MB'' water model in
Ref.~\cite{Dillea98}, where  an average  strengthening of shell 
hydrogen bonds is detected, too, as well as negative values of $\Delta
C$ at low temperatures. 
 
It should be mentioned, though, that negative $\Delta E$ and $\Delta
S$, and positive $\Delta F$ can be obtained for the MB model within the 
information theory approximation applied to the cavity opening in
pure water Ref.~\cite{ArtHay99}, even if, as already stated, this
approach cannot recover the correct trends for the specific heat. 
Since cavity opening disregards 
bond differences in bulk and
shell, this fact would imply that hydrogen bond differences are irrelevant
to the subject. 

It is likely that both cavity formation and  hydrogen-bond 
differences contribute to the hydrophobic hydration, and an ultimate
answer  will come from simulations in three dimensions when they will  
become accurate enough as to predict specific heats.
As far as the present model is concerned, we consider the good overall
agreement with experiments and two-dimensional simulations as
satisfactory to justify the use of this model in polymer hydration
studies. 

\subsection{Cold and warm unfolding of a non-polar homopolymer}

\begin{figure}
\epsfxsize= 8 truecm 
\begin{center}
\mbox{\epsfbox{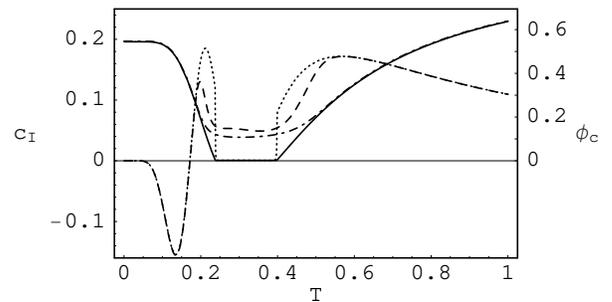}} 
\end{center}
\caption{Nonpolar homopolymer in solution: temperature profiles, 
for different lengths $N$, of the
excess specific heat per water site, 
$c_I=C_I/N_W$ (left scale; dashed line: $N=200$; dotted,
$N=10^6$), and 
average fraction of water-monomer contacts,
$\phi_c= \langle n_c \rangle/(2 N + 2)$ (right scale; 
dash-dotted line: $N=200$; solid,
$N=10^6$).
Parameters as before.
}
\label{fig:polymer}
\end{figure}

Let us now turn to the study of a nonpolar 
homopolymer in solution, taking into account 
just the  behavior of water clusters in  the vicinity 
of a monomer, and disregarding interactions between monomers 
and  between water clusters. Our goal is, in fact, to understand 
the effect of the hydrophobic effect alone on polymer  behavior.
We model a polymer as a $N$-step 
self-avoiding walk (SAW) on a two dimensional lattice with coordination $z$.
On each lattice site there can be
either a monomer or $z$ clusters of $m$ water molecules, so that each 
monomer-water contact involves one cluster. 
The Hamiltonian follows from Eq.\ (\ref{hamclu}):
\beq
{\cal H} = \sum_{j=1}^{N_W}  \sum_{\mu=1}^{z} 
\varepsilon_{j \mu} \, + K m l_j \,,
\label{hampoly}
\eeq
where $N_W= (z-2) N + 2$ is the  highest number of water sites that can 
be in contact with the polymer, and $l_j$ is the number of contacts 
between the $j$-th water site and the monomers.
The partition function of the polymer in solution reads as:
\beq
{\cal Z} = \sum_{C} Z(C) = \sum_{n_c}
\zeta \left(n_c \right) Z (n_c)
 \eeq
where $C$ are the conformations of the polymer and $Z(C)$ 
the restricted partition function, obtained tracing over water variables
at fixed conformation $C$. Due to the form of the Hamiltonian given in 
Eq.\ (\ref{hampoly}),  $Z(C)$ depends only on the total number of 
water-monomer contacts $n_c$, and $\zeta \left(n_c \right)$ is the 
number of  SAWs  characterized by the same value of $n_c$.
$\cal Z$ can be factorized as
${\cal Z} = {\cal Z}_b {\cal Z}_I$, 
where ${\cal Z}_b =({\cal Z}_b^{\text{clu}})^{ z N_W}= 
A_b^{z N_W}$ is the contribution of all water sites when in
contact to other water, and ${\cal  Z}_I = \sum_{n_c} \zeta \left( n_c \right)
x^{n_c}$.
 According to the above factorization, the specific heat 
 is the sum of a bulk contribution
$C_b$ and of an interaction contribution $C_I$. We introduce also
the average number of  water-monomer contacts,
$\langle n_c \rangle = x \frac{\partial}{\partial x } \log{ {\cal Z}_I}$,
which is a measure of the compactness of the polymer.
To exactly evaluate the above  quantities, $\zeta \left( n_c \right)$ should be
obtained through an exhaustive enumeration of 
the SAWs. 
However, if we restrict ourselves to a square lattice, the numerical
estimates reported in \cite{Dougea} allow us to write
\beq
\zeta \left( n_c \right) \simeq \zeta_0 \frac{1}{w(n_c)!} 
(\alpha_0 N)^{w(n_c)} \exp(\alpha_0 N)~,
\eeq
where $w(n_c)=(N_W-n_c)/2$ is the number of monomer-monomer contacts, 
$\alpha_0=0.164$ and $\zeta_0$ is the  number of SAWs of length $N$. 
Such an
estimate is expected to be very good if $N$ is large \cite{Dougea}.
Hence, an analytical expression can be found for ${\cal Z}_I$:
\beq
{\cal Z}_I=\frac{1}{w_{\text{max}}!}x^{N_W} 
\Gamma (w_{\text{max}}+1, \frac{\alpha_0 N}{x^2}) 
e^{  -\alpha_0 N \left( 1- \frac{1}{x^2} \right) } \,,
\label {zian}
\eeq
where $\Gamma(n,x)$ is the incomplete $\Gamma$-function and 
$ w_{\text{max} } =(N_W - n_c^{\text{min} })/2 $
(the smallest number of contacts is the perimeter of the globule, 
$n_c^{\text{min}} \simeq 2 \sqrt{N \pi}$).
The results for $C_I$ and  $\langle n_c \rangle$ 
are reported in Fig.\ \ref{fig:polymer}.

The presence of both ``cold'' and ``warm'' collapse transitions, signalled by
the drop of the number of contacts and by the jumps in the specific heat is 
strikingly evident.   
Notice that the polymer swells at low temperature even with $K=0$: 
water-clusters' energy at $T=0$ is not affected by the contact with  monomers, 
and cold unfolding  is not a trivial 
effect of  a  swollen-ground-state  dominance.  
As $N$ grows both transitions get sharper,
thus suggesting the existence of true phase transitions in the thermodynamic
limit. 
The above phenomenology is very close to that of proteins in
solution \cite{Privalov}. 
Moreover,  swelling at low temperatures has  been found in homopolymers
like poly(N-isopropylacrylamide) (PNIPAM) and
poly(N-isopropylmethacrylamide) (PNIPMAM) \cite{pnipam}, while 
warm swelling  has been detected  for a number of homopolymers 
in non-aqueous solution \cite{heat}.   

Both PNIPAM and PNIPMAM present a polar peptide group in the
side chain and hence are able to form hydrogen bonds; 
hence, one could object that it is not appropriate to propose a
comparison with  their phenomenology, since polar groups favourably
interact with water and help the swelling of the chain.
Actually, we are not aware
of any calorimetric experiment on completely nonpolar homopolymers in
water (that would represent ideal tests to  our predictions, provided
that they swell in the range $T= 0-100\, ^{\circ} C$ 
and the solution is so dilute 
that aggregation is negligible).
However,  hydrogen bonds can be formed both in the coil phase, with
water,  and in the
globule one (and are indeed considered a possible cause for the detected 
hysteresis in the coil-globule transition \cite{Wangea98}).  Assuming 
a negligible overall
contribution of hydrogen bonds to the stability of the coil over the
globule phase, one can  admit that the cold transition in
PNIPAM and PNIPMAM is mainly due to the temperature dependence of the
free-energy of hydration of the nonpolar groups.
This is why we think that it is meaningful to compare our prediction
for a nonpolar homopolymer to experimental results on PNIPAM and PNIPMAM. Our
model is not intended to give a detailed description of their
phenomenology, though: because of its dimensionality, of
its lack of geometric detail, and of the assumptions that make it
simple, it is not suited to describe the ``domain'' cooperativity 
within of the coil-globule
transition, or the freezing of side chains within the globule. 
 
Warm swelling of PNIPAM and PNIPMAM has not been detected: this is due to
the fact that the temperature of warm transition should lie  above
the water boiling temperature  \cite{Graz2000}. 

\section{Concluding remarks}

We have presented a simple model which 
reproduces the two-dimensional simulation 
\cite{Dillea98} and the 
experimental results \cite{Privalov} for the  trends
of $\Delta F$, $\Delta E$, $\Delta S$, $\Delta C$  upon hydration of a
nonpolar solute, showing that the reduced number of ways of forming
hydrogen bonds in the presence of a non-bonding solute, together with a
shift in hydrogen bonding energy, are sufficient to reproduce the
experimental hallmarks of hydrophobic hydration.  This is consistent with
the views expressed in  Refs.~\cite{Dillea99,Dillea98}  and
\cite{Matu94}.   

Our description of  water is focused on the clusters in the hydration
shell, which are affected by the presence  of the solute molecule. To
keep the model  analitically integrable, we disregard interaction
between clusters. In this way, we are able to describe  hydration, but 
cannot describe solvent-mediated long-range interactions between
solute molecules. For this reason, 
this approach is somewhat complementary to the one recently
proposed by Kolomeisky and Widom \cite{KoWi}, which can be treated
analytically in the one-dimensional case, and  is very well
suited to evaluate the potential of mean force.

When applied to a homopolymer in solution, our model recovers the
experimental swelling at both high and low temperatures. We
reported only a calculation for a two-dimensional  polymer. We did so for
the sake of the internal coherence of the model,  because the geometric
contribution to the density of states of water clusters is estimated for
the two-dimensional case and the entropies of both the water and the
polymer play a key role in the behavior of the polymer in solution. 
Anyway, the fact that our two-dimensional water yields a correct
qualitative  description of nonpolar hydration suggests that the
application of the 2-$d$ water model to the hydration of 3-$d$ SAWs could
be interesting; indeed, we have performed some calculations using
three-dimensional SAWs and have verified that they also reproduce the same
qualitative results; of course, the temperature range of the compact phase
varies a bit.

The fact that our model describes both cold and warm swelling of a
nonpolar homopolymer suggests that it could be effectively used in protein
folding studies, describing both cold and warm unfolding in a single
framework.     Even if hydrophobicity alone cannot be safely considered as
the unique responsible  for protein folding and stability, it is very
likely that any progress in its description,  and in its encoding in
simple models, will yield significative improvement in the understanding
of the protein folding puzzle. 

\acknowledgments

We thank S.~Causo, P.~De Los Rios and  A.~Pelizzola for very
useful discussions and suggestions. A large part of the present work was
carried out when one of the authors (LC) was at the INFM, Politecnico di
Torino.

\end{multicols}

\begin{references}
\bibitem{note_gamma} Assuming e.g.\ that the bond is lost in the
``I,E'' (rather than in the ``E,I'') disposition,  there are
${{s} \choose {k}}$ ways of choosing the ``I''s and ${{m-s} \choose {k}}$ of
choosing the ``E''s for the $k$ ``I,E'' interfaces.
\end{references}
\end{document}